\documentclass[aps,twocolumn,showkeys,groupedaddress]{revtex4-1}
\usepackage{amsmath,amsthm,verbatim,amssymb,amsfonts,amscd,relsize}
\usepackage{graphicx}
\usepackage{gensymb}
\usepackage{booktabs}
\usepackage{xcolor}
\usepackage{epstopdf}
\begin{document}

\title{Quasichemical theory and the description of associating fluids relative to a reference: Multiple bonding of a single site solute}

\author{Artee Bansal}
\affiliation{Chemical and Biomolecular Engineering, Rice University, Houston, TX}
\author{Walter G. Chapman}
\affiliation{Chemical and Biomolecular Engineering, Rice University, Houston, TX}
\author{D. Asthagiri}\thanks{Email: dna6@rice.edu}
\affiliation{Chemical and Biomolecular Engineering, Rice University, Houston, TX}

\date{\today}

\begin{abstract}
We derive an expression for the chemical potential of an associating solute in a solvent relative to the value in a reference
fluid using the quasichemical organization of the potential distribution theorem. The fraction of times the solute is not associated with the solvent, the monomer fraction, is expressed in terms of (a) the statistics of occupancy of the solvent around the solute in the reference fluid and (b) the Widom factors that arise because of turning on solute-solvent association.  Assuming pair-additivity, we expand the Widom factor into a product of Mayer $f$-functions and the resulting expression is rearranged to reveal a form of the monomer fraction that is analogous to that used within the statistical associating fluid theory (SAFT).  The present formulation avoids all graph-theoretic arguments
and provides a fresh, more intuitive, perspective on Wertheim's theory and SAFT. Importantly, multi-body effects are transparently incorporated into the very foundations of the theory.  We illustrate the generality of the present approach by considering examples of multiple solvent association to a colloid solute with bonding domains that range from a small patch on the sphere, a Janus particle, and a solute whose entire surface is available for association. 
 \end{abstract}
\maketitle

\section{Introduction}

The statistical associating fluid theory (SAFT) \cite{chapman_new_1990,chapman_phase_1988} is a well-established framework for modeling associating fluids. 
SAFT, and Wertheim's theory \cite{wertheim_fluids_1984,wertheim_fluids_1984-1,wertheim_fluids_1986,wertheim_fluids_1986-1} on which it is based, seek to describe the physics of short-range association given the properties of a reference fluid, typically a hard-sphere or Lennard-Jones fluid. The associating solute and solvent are envisioned as having sticky-patches on their surfaces over which they bond; the range of attraction is very short relative to the size of the particle.  The Helmholtz free energy of this associating fluid is then constructed using as key ingredients the pair-correlation information from the reference, the first order thermodynamic perturbation theory (TPT1), or triplet correlation that go beyond the superposition approximation, the second order perturbation theory (TPT2) \cite{marshall_wertheims_2012,marshall_thermodynamic_2014}. TPT1 works well when the attraction is of short-range and the sticky patches are restricted to bond only once. TPT2 works well for systems where patches are restricted to bound a maximum of two particles \cite{marshall_wertheims_2012,marshall_thermodynamic_2014}. However TPT2 fails for patch geometries that allow bonding more than two times.   But for many problems, such as those involving colloidal-solvent or ion-solvent association, acknowledging multiple bonding ($>2$) at a site, and multi-body effects in general, becomes essential. 

Earlier Marshall and Chapman \cite{marshall_molecular_2013,marshall_thermodynamic_2013} had suggested an approach to model multiple solvent bonding to a spherically symmetric association site within the TPT2 framework. 
In essence they approximate the integral 
 \begin{eqnarray*} 
\int_{v} \ldots \int_{v} d\vec r_1 \cdots d\vec r_n\, g_{HS}(\vec r_1 \cdots \vec r_n |0) ,
\end{eqnarray*}
of  the $n$-particle correlation $g_{HS}(\vec r_1 \cdots \vec r_n |0)$ of the hard-sphere solvent given the hard-sphere solute  at the origin $(\ldots | 0)$ by  $y_{HS}^n(d) \delta ^{(n)} \Xi ^{(n)}$, where $y_{HS}(d)$ is the cavity correlation function at contact for particles of diameter $d$, $\delta$ is a correction to account for three body interactions, and $\Xi ^{(n)}$ is related to the cluster integral of $n$-solvent plus one solute particle within the observation volume $v$. The cluster integrals are calculated separately by a Monte Carlo procedure. 

The Marshall-Chapman approach \cite{marshall_molecular_2013,marshall_thermodynamic_2013} works quite well for low particle densities, typically $\rho d^3 \leq 0.6$, but deviations in structural and thermodynamic properties become serious for higher densities and/or bonding strengths. Building on the Marshall-Chapman approach, and  drawing inspiration from the quasichemical theory of solutions \cite{lrp:apc02,lrp:book,lrp:cpms}, earlier we \cite{bansal_structure_2016,bansal_thermodynamics_2017} developed an approach to model the above integral by the occupancy distribution of the solvent around the solute within the observation volume $v$. Importantly in our approach the occupancy 
distribution was obtained from the reference fluid at the same density as the solution being modeled.  This complete reference approach \cite{bansal_structure_2016,bansal_thermodynamics_2017} is able to describe accurately  the structure and thermodynamics of a colloidal solute that can bond multiple patchy-solvents for a range of system densities, bonding energies, and solute-solvent size asymmetries. 

The above development motivated us to reconsider the problem of modeling associating fluids entirely within the quasichemical (QC) organization of the potential distribution theorem (PDT) \cite{lrp:apc02,lrp:book,lrp:cpms}. As has been emphasized earlier \cite{lrp:book,lrp:cpms}, the potential distribution theorem provides a general basis for the theory of solutions and a tool to develop physically motivated approximate models of solution thermodynamics.  The potential distribution theorem presents a \emph{local} partition function to be evaluated for the excess chemical potential of the defined component.  Importantly, this partition function can be recast as a summation over physical clusters within the defined observation volume, leading to the quasichemical organization of the potential distribution theorem. The equilibrium constant for forming the clusters then plays an important role in the theory. While the calculation of the equilibrium constant is not trivial for most problem 
 of interest, a definite virtue of the approach is its 
rather clear physical underpinnings. Importantly, the many-body aspect of clustering is built into the very foundations of the theory. 

 It is helpful to contrast the QC/PDT approach with the Wertheim/SAFT approach. The latter rests on an expansion of the grand-potential of the \emph{entire} system in terms of physically bonded components, but to discover the physically bonded clusters requires use of subtle graph-theoretic ideas. A virtue of the Wertheim/SAFT approach is that it makes available the excess Helmholtz free energy of the system, making its use in applications easier. But, as noted above, incorporating many-body correlations in Wertheim/SAFT is not as transparent as in QC/PDT.  For problems requiring attention to many-body correlations, QC/PDT may helpfully complement Wertheim/SAFT. With this broader goal, here we present the key steps for describing associating fluids within QC/PDT.

We explore a range of bonding configurations from a solute that can bond only once, to a solute that can bond multiple solvents but only on one-hemisphere of its surface, i.e.\ a Janus particle \cite{roh_biphasic_2005,walther_janus_2013}, and to a solute with a sticky patch that covers its entire surface. Quasichemical theory leads to the
identification of the occupancy of a patch conditional on the total occupancy of the observation volume, all in the reference fluid, as an important quantity within the theory. We suggest a Monte Carlo procedure to calculate this for general cases and provide analytical models for limiting cases of a patch that can bond only once or a patch that covers the entire surface of the solute. For simplicity, in the present work we consider a system where only solute-solvent bonding is allowed. Forthcoming studies will ease this requirement. 

The rest of the article is organized as follows. In Section~\ref{sc:theory} we sketch the quasichemical approach and then develop the idea of association relative to a non-associating reference.  In Section~\ref{sc:methods} we present the methods, and in Section~\ref{sc:results} we present the results from several model systems. 

\section{Quasichemical theory}\label{sc:theory}

The excess chemical potential, $\mu^{\rm ex}$, of a solute in a solvent is that part of the Gibbs free energy of solvation that arises due to intermolecular interactions; $\mu^{\rm ex}$ is defined relative to the ideal gas at the same density and temperature. Formally, $\mu^{\rm ex}$ is given by the potential distribution relation
\begin{eqnarray}
\beta \mu^{\rm ex} = \ln \int e^{\beta \varepsilon} P(\varepsilon) d\varepsilon \, ,
\label{eq:pdt}
\end{eqnarray}
where $P(\varepsilon)$ is the probability density distribution of the solute-solvent binding energies, and as usual, $\beta = 1/k_{\rm B}T$. 

\begin{figure*}[ht!]
\begin{center}
\includegraphics{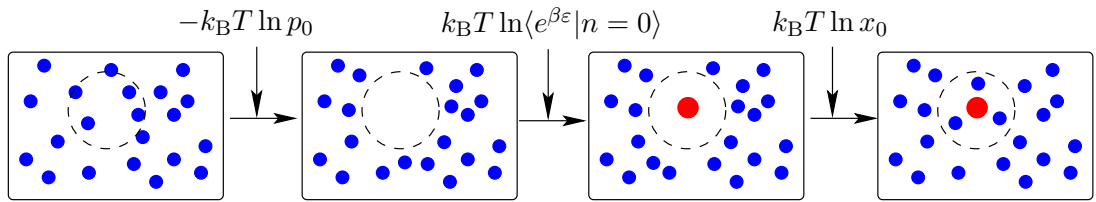}
\end{center}
\caption{Schematic of the quasichemical organization of Eq.~\ref{eq:pdt}. 
Adapted from Figure 1, J.\ Chem.\ Phys.\ {\bf 130}, 195102 (2009).}
\label{fg:qcpdt}
\end{figure*}

A direct application of Eq.~\ref{eq:pdt} is almost never satisfactory because the high-$\varepsilon$ tail of the probability distribution, which reflects short-range repulsive interactions, is usually difficult to characterize.  In the quasichemical (QC) approach \cite{lrp:book,lrp:cpms}, we separate the short-range and long-range contributions and include the contribution from the short-range solute-solvent interactions within a chemical equilibrium framework.  To this end
we demarcate a domain, the observation volume or inner-shell, around the solute (Fig.~\ref{fg:qcpdt}). 
The observation volume is usually restricted to the first solvation shell of the solute. The probability of observing $n$-solvent particles within the observation volume is denoted as $x_n$.  The reversible work to empty the observation volume is $-\beta^{-1}\ln x_0$. The solute with an empty observation shell, the renormalized solute, interacts with the bulk fluid solely through long-range interactions. In particular, for a sufficiently large inner-shell radius, the solute-solvent 
binding energy of the renormalized solute is Gaussian.  To complete the thermodynamic description of solvation, we also need to account for the free energy to create a bare cavity of the size and shape of the observation volume. This free energy is denoted as $-\beta^{-1} \ln p_0$, where $p_0$ is the probability to form an empty cavity in the bulk fluid. Thus the excess chemical potential is written as 
\begin{eqnarray}
\beta \mu^{\rm ex} = \ln x_0 - \ln p_0 + \ln \langle e^{\beta \varepsilon} | n = 0\rangle \, .
\label{eq:pdt1}
\end{eqnarray}
Fig.~\ref{fg:qcpdt} provides a schematic of the quasichemical organization. As discussed next, the appellation quasichemical derives from the chemical organization of $x_0$ and $p_0$. 

Consider the chemical equilibrium between the solute ($\sigma$) and solvent ($s$)  to form an $n$-solvent cluster, 
\begin{eqnarray*}
\sigma + n\cdot s \rightleftharpoons \sigma s_n
\end{eqnarray*}
The usual products-over-reactants equilibrium constant is given by $K_n = x_n/x_0 \rho_s^n$, where $\rho_s$ is the density of the solvent. A mass balance \cite{lrp:book,lrp:cpms,merchant_thermodynamically_2009} then gives 
\begin{eqnarray}
\ln x_0 = -\ln\left[1 + \sum_{n \geq 1} K_n\rho_s^n\right] \, .
\label{eq:x0}
\end{eqnarray}
A similar equation can be written for $p_0$, 
 \begin{eqnarray}
\ln p_0 = -\ln\left[1 + \sum_{n \geq 1} \tilde{K}_n\rho_s^n\right] \, .
\end{eqnarray}
where $\tilde{K}_n$ is the equilibrium constant for the association between a bare cavity and $n$-solvent particles. 

$K_n$ is related to configurational integrals by 
\begin{equation}
K_n = \frac{(e^{ \beta\mu^{\rm ex}_{s} })^n}{n!} e^{-\beta w(R^n)} \int\limits_vd{\vec r}_1\ldots\int\limits_v d{\vec r}_n \; e^{-\beta U_{\sigma s_n}(R^n)} \, ,
\label{eq:kn}
\end{equation}
where $v$ is the volume of the inner-shell, $U_{\sigma s_n}(R^n)$ is the potential energy of the solute-$n$-solvent cluster, $\mu^{\rm ex}_s$ is the excess chemical potential of the solvent, and $e^{-\beta w(R^n)}= \langle e^{-\beta \phi(R^n;\beta)} | R^n \rangle_0$ is the average of the interaction free energy between the cluster and the bulk. Here $\langle \ldots| R^n \rangle_0$ indicates averaging  over the normalized probability density for cluster conformations $R^n$ in the absence of interactions with the rest of the medium.

\subsection{Quasichemical perspective of associating fluids}\label{sc:QCA}
Associating fluids are characterized by short range, directional interactions.  In SAFT and in Wertheim's theory \cite{muller_molecular-based_2001,chapman_phase_1988,wertheim_fluids_1984,chapman_new_1990,wertheim_fluids_1984-1}, we assume the availability of a well-characterized reference and the role of association is considered relative to the reference. For definiteness, we assume the reference is a hard-sphere fluid. The above quasichemical organization can be used for the reference as well, and we distinguish all the reference properties with the subscript $r$. 

We first focus on $x_0$ and rewrite this in terms of the properties of the reference. From Eqs.~\ref{eq:x0} and~\ref{eq:kn} and the corresponding relations for the reference, 
we can re-express the chemistry contribution relative to the reference as
\begin{widetext}
\begin{eqnarray}
\ln x_0 = -\ln\left[1 + \sum_{n \geq 1} \frac{K_n}{K_{n,r}} K_{n,r}\rho_s^n\right] = \ln x_{0,r} - \ln\left[x_{0,r} + \sum_{n\geq 1} \frac{K_n}{K_{n,r}}  x_{n,r}\right] \, 
\label{eq:widomSmall}
\end{eqnarray}
\end{widetext}
Let us next examine the ratio of the equilibrium constants. We have 
\begin{eqnarray}
\frac{K_n}{K_{n,r}} = e^{n\cdot \beta\Delta \mu^{\rm ex}_s} \cdot e^{\beta\Delta w(R^n)} \cdot \langle e^{-\beta \Delta \tilde{U}_{\sigma s_n}(R^n)}| R^n \rangle_{r} \, 
\label{eq:Knratio}
\end{eqnarray}
where $\Delta \tilde{U}_{\sigma s_n}(R^n)$ is the potential energy of the $n$-solvent plus solute cluster in the physical system relative to the reference. Likewise,  
$\Delta \mu^{\rm ex}_s$ and $\Delta w(R^n)$ are the corresponding properties relative to the reference. 
The factor $e^{n\cdot \beta\Delta \mu^{\rm ex}_s} \cdot e^{\beta\Delta w(R^n)}$ accounts for the entropic effects in sequestering the solvent within the observation volume above the effect in the reference system. The factor $\langle e^{-\beta \Delta \tilde{U}_{\sigma s_n}(R^n)} | R^n \rangle_{r}$ are Widom-factors, but now relative to a reference and 
in the volume $v$. The above relations are exact within classical statistical mechanics for any arbitrary forcefield. Appendix (Section~ \ref{sc:appen}) provides a concise
derivation of Eq.~\ref{eq:Knratio}.
% We next simplify these relations for pair-additive forcefields and the usual bonding situation addressed within Wertheim/SAFT. 

The potential energy $\Delta \tilde{U}_{\sigma s_n}$ can be partitioned into solvent-solvent ($\Delta U_{ss}$) and solute-solvent ($\Delta U_{\sigma s_n}$) contributions. The presence of solvent-solvent short-range interaction can be incorporated by factoring the Widom factor as $\langle e^{-\beta \Delta U_{ss}} | R^n\rangle_{r} \cdot 
\langle e^{-\beta \Delta U_{\sigma s_n}} | R^n\rangle_{r+ss}$, where $\langle\ldots \rangle_{r+ss}$ indicates averaging over the case where solute-solvent interactions are
reference interactions and solvent-solvent interactions including short-range bonding.  (There are other ways to incorporate solvent-solvent association effects, but we will not consider those in the present paper.)

\begin{figure*}[!ht]
	\begin{center}
		\includegraphics[scale=0.35]{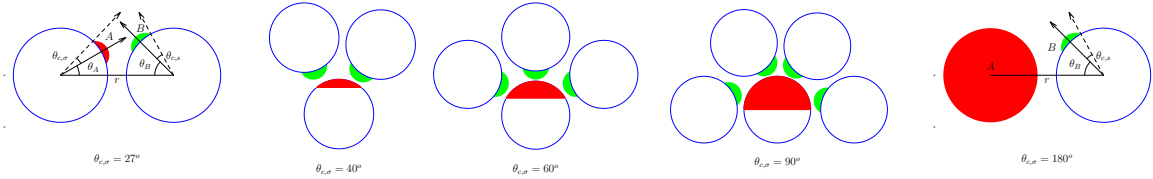}
		\caption{Different patch geometries ($\theta_{c,\sigma}$) of the solute molecule. For a solute with  ($\theta_{c,\sigma}=180^{\degree}$), the attractive patch A can be approached from any direction. Note that the critical  angle for the solvent molecules $\theta_{c,s}=27\degree$, so that the patch $B$(colored green) on the solvent can only interact once with the multi-bonding patch $A$(colored red) on the solute molecule. It should be noted that due to the 2-D nature of the graph, all the possible bonding conditions can not be represented. Table~\ref{tab:diff_patch} gives the maximum bonding numbers for different patch sizes on the solute molecule. }
		\label{fg:diff_patch}
	\end{center}
\end{figure*}

We consider solvent bonding configurations  such that solvent-solvent bonding within the solute's inner-shell is avoided. In this case, $\langle e^{-\beta \Delta U_{ss}} | R^n\rangle_{r} = 1$ and $\langle e^{-\beta \Delta U_{\sigma s_n}} | R^n\rangle_{r+ss} = \langle e^{-\beta \Delta U_{\sigma s_n}} | R^n\rangle_{r}$.
Further, for a pair additive forcefield, $\Delta U_{\sigma s_n}(R^n) = \sum_{i=1}^n \Delta U_{\sigma s_i}$. Thus 
\begin{widetext}
\begin{eqnarray}
e^{-\beta \Delta U_{\sigma s_n}(R^n)} = \prod_{i=1}^n (1+f_{\sigma i}) = 1 + \sum_i f_{\sigma i} + \sum_{1 \leq k < j \leq n} f_{\sigma k}f_{\sigma j} + \ldots ,
\label{eq:mayer1}
\end{eqnarray}
\end{widetext}
where $f_{\sigma i} = e^{-\beta \Delta U_{\sigma i}} - 1$ is the Mayer $f$-function for association between the solute and the $i^{\rm th}$ solvent within the observation volume.  Each term in the above expansion is a contribution due to association when one, two, $\ldots$, $n$ solvent particles bond with the solute, given that $n$ solvent 
particles in the inner shell of the solute. As is typically assumed in SAFT and Wertheim's approach, we assume the association strength is the same for all the solvent-solute pairs. The association potential \cite{bol_MontCarlo_1982} for the solute-solvent pair $(\sigma,s)$ is given by:
\begin{equation}
\Delta u_{(\sigma,s)}^{AB}{(r)}=
\begin{cases}
-\epsilon ~, r<r_c \,\text{and}\, \theta_A\leq \theta_{c,\sigma}^{(A)}\,\text{and}\,\theta_B \leq \theta_{c,s}^{(B)}
\\
0   \text{ \ \ \ \ \ otherwise}
\\    
\end{cases}
\label{eq:potential1}
\end{equation}
where the superscripts $A$ and $B$ represent the type of site on, respectively, the solute and the solvent, and $\epsilon$  is the association energy; $f_{\sigma} = e^{-\beta \epsilon} - 1$ is the corresponding Mayer $f$ function. $r$ is the distance between the particles and $\theta_A$ is the angle between the vector connecting the centers of two molecules and the vector connecting association site $A$ to the center of that molecule (Fig.~\ref{fg:diff_patch}).  The critical distance beyond which particles do not interact is $r_c$. The angular extents of the patch for the solute and the solvent molecules are, respectively, $\theta_{c,\sigma}^{(A)}$ and  $\theta_{c,s}^{(B)}$; if the inter-particle vector falls outside a patch, the particles cannot bond. 
The angular span of the patch $\theta_{c,\sigma}^{(A)}$ on the solute molecule determines whether the solute can bond one or more solvent particles (Fig.~\ref{fg:diff_patch}). 

Denoting the Mayer $f$-functions as $f_\sigma$, we thus find 
\begin{eqnarray}
\langle e^{-\beta \Delta U_{\sigma s_n}(R^n)} | R^n \rangle_{r} = 1 & + & \sum_{1 \leq i\leq n}\langle  f_{\sigma}|R^n \rangle_{r} \nonumber \\
& + & \sum_{1 \leq k < j \leq n} \langle  f_{\sigma }^2|R^n \rangle_{r} \nonumber \\
& + & \ldots 
\label{eq:ensm_avg}
\end{eqnarray}

Consider the term $\langle  f_{\sigma}|R^n\rangle_{r}$. This is the average Mayer $f$-function for a single solvent particle interacting with the solute. 
Since there are $n$-solvent particles in the observation volume, there are $n$ choices of particles to occupy the bonding patch (solid bond angle) on the solute, leading
to the summation  $\sum_{1 \leq i\leq n}$. The other terms in the expansion can be given a similar meaning.  It proves helpful to codify the combinatorics
by means of an indicator function. Specifically, we define 
\begin{eqnarray}
\chi(i, n)= \begin{cases}
1, & \text{if at least } i \leq n \; \text{solvent} \\
     & \text{occupy the bonding patch} \\
 0, & \text{otherwise} \\
 \end{cases}
\label{eq:combinations}
\end{eqnarray}
\begin{widetext}
The $(i+1)^{\rm th}$ term on the right hand side of Eq.~\ref{eq:ensm_avg} is then 
\begin{eqnarray}
\sum_{1 \leq a < \ldots < i \leq n}  \langle f_{\sigma }^i |R^n \rangle_{r}  \equiv    \langle f_{\sigma}^i \chi(i, n)|R^n\rangle_{r} = \frac{ \langle f_{\sigma}^i \chi(i,n)|R^n\rangle_{r} }{ \langle \chi(i,n)|R^n\rangle_{r} } \langle \chi(i,n)|R^n\rangle_{r} =  \langle f_{\sigma}^i | i \cdot R^n \rangle_{r}\cdot  \langle \chi(i,n) |R^n\rangle_{r}
\label{eq:fi_avg}
\end{eqnarray}
\end{widetext}
In deriving the last term in Eq.~\ref{eq:fi_avg} we have used the rule-of-averages \cite{lrp:apc02,lrp:book,lrp:cpms}, and $\langle\ldots | i \cdot R^n \rangle_{r}$ indicates that  (a) $n$-solvent particles are in the inner shell \textbf{and} (b) \emph{at least} $i$ of the $n$ solvent particles are over the bonding patch. When the association strength is the same for all the solute-solvent pairs, as in the usual cases in  Wertheim/SAFT, the solvent on the solute patch bonds independently of the bonding state of the other solvent particles, we find
\begin{eqnarray}
\langle f_{\sigma}^i | i \cdot R^n \rangle_{r}  = \bar{f}_\sigma^i \cdot \kappa_s^i \, ,
\end{eqnarray}
where $\kappa_s = [1-\cos(\theta_{c,s})]/2$  is the probability that the solvent molecule is oriented such that it can bond with the solute and $\bar{f}$ accounts
for averaging over the radial coordinate. For a square-well potential (Eq.~\ref{eq:potential1}), we simply have  $f = \bar{f} = e^{-\beta \varepsilon} - 1$. 
The factor $\langle \chi(i)|R^n \rangle [= Q(i|n)]$ is the probability that given $n$-solvent particles in the inner-shell, \emph{at least} $i \leq n$ occupy the patch region and are
available to bond with the solute. 

Putting all of the above together, we have 
 \begin{eqnarray}
\langle e^{-\beta \Delta U_{\sigma s_n}(R^n)} | R^n \rangle_{r}  & = &  1 + \nonumber \\ 
& & \sum_{1\leq i \leq n}   \bar{f}_{\sigma}^i\cdot \kappa_s^i \cdot Q(i|n) 
 \label{eq:Q_exp1}
 \end{eqnarray}
 
 Going back to Eq.~\ref{eq:Knratio}, we provisionally assume that the surface term is also pair-decomposable, with each pair-wise contribution the same. 
 (We emphasize that pair decomposability is assumed for a free energy rather than an interaction potential, but this assumption can be relaxed.) Thus, provisionally we set $\beta\Delta w(R^n) = n\cdot \beta \Delta w$. For convenience we write $\xi_s = \exp({\beta\Delta\mu^{\rm ex}_s})\exp({\beta\Delta w})$. 
  
Thus we finally obtain 
\begin{eqnarray}
\frac{K_n}{K_{n,r}} = \xi_s^n \cdot    \left[1+ \sum_{1\leq i \leq n} \bar{f}_{\sigma}^i\cdot \kappa_s^i \cdot Q(i|n)  \right]\, 
\label{eq:Kn_ratio_Q}
\end{eqnarray}  

Substituting the above ratio in Eq.~\ref{eq:widomSmall}, we find
\begin{widetext}
\begin{eqnarray}
\ln x_0 &=& \ln x_{0,r} - \ln\left[  x_{0,r} + \sum_{n\geq 1} \xi_s^n \cdot    \left[1+\sum_{1\leq i \leq n}  \bar{f}_{\sigma}^i\cdot \kappa_s^i \cdot Q(i|n) \right] \cdot x_{n,r} \right ] \nonumber \\ & =& \ln x_{0,r} + \ln X_\sigma \, . 
\label{eq:widomSaft}
\end{eqnarray}
\end{widetext}
The physical meaning of the above equation is the following. The chemical work (Fig.~\ref{fg:qcpdt}) is composed in two steps: (1) the free energy $\ln x_{0,r}$
to populate the observation shell with reference solvent particles, and (2) the free energy $\ln X_\sigma$ to turn on solute-solvent  and solvent-solvent association. 
Eq.~\ref{eq:widomSaft} is the principal contribution of this work. 

We can pursue a similar development for the packing contribution (Fig.~\ref{fg:qcpdt}).  The final form of the association contribution to the 
excess chemical potential of the solute is thus 
\begin{eqnarray}
\mu^{\rm ex}_{\rm asso} & =& \ln \frac{x_{0}}{x_{0,r}} -\ln \frac{p_{0}}{p_{0,r}} \nonumber \\
& = &  \ln X_\sigma - \ln P_{\sigma} \, ,  
\label{eq:brk_cpt}
\end{eqnarray}
where the $\ln P_\sigma$ term arises solely from association contribution to solvent reorganization.

Given the ratio of the equilibrium constants (${K_n}/{K_{n,r}}$) and $X_\sigma$, the occupancy distribution of solvent molecules around the associating solute is
\begin{eqnarray}
x_n &=& K_n \cdot \rho_s^n\cdot x_0 \nonumber \\ 
& =& \frac {K_n}{K_{n,r}}  \cdot x_{n,r}\cdot X_\sigma
\label{eq:xn_asso}
\end{eqnarray}

The bonding distribution can be obtained from this occupancy distribution using standard rules of probability
\begin{eqnarray}
X_i =  \sum_{n \geq i}x_n \cdot P(X_i | n),\    \forall \, n\geq 0 \, ,
\label{eq:Xn}
\end{eqnarray}
where $P(X_i | n)$ is the conditional probability of having $i$ bonded solvents when $n$ solvent molecules are present in the observation volume. 
Note that 
\begin{eqnarray}
\sum_{i \leq n}P(X_i | n) = 1, \, \forall\, n \geq 0
\label{eq:PXnconsistency}
\end{eqnarray}

\subsection{Single site solute}\label{sc:QCA_noPP}

To maximize clarity and simplify the analysis of the association contribution, in this article we consider the case of an infinitely dilute solute in a solvent which can not associate with other solvent molecules, i.e.\ 
only solute-solvent association is allowed. For this case  $\xi_s=1$; please note that there will also be some contribution to the chemical potential of the solvent due to solute-solvent association, but for an infinitely dilute case this can be neglected. The equilibrium ratio simplifies to
\begin{eqnarray}
\frac{K_n}{K_{n,r}} =  \left[1+ \sum_{1\leq i \leq n} \bar{f}_{\sigma}^i\cdot \kappa_s^i \cdot Q(i|n)  \right]\, \, .
\label{eq:Kn_ratio_Q_red}
\end{eqnarray}
For this system, $X_\sigma$ is equal to $X_0$, the fraction of times the solute is not bonded, i.e.\ the monomer fraction, a quantity that
plays a central role in Wertheim/SAFT. For the chemical contribution, expanding and rearranging the order of summation in the second term on the right hand side of Eq.~\ref{eq:widomSaft}, we have 
\begin{eqnarray}
x_{0,r} + \sum_{n\geq 1}     \left[1+\sum_{1\leq i \leq n}  \bar{f}_{\sigma}^i\cdot \kappa_s^i \cdot Q(i|n) \right] \cdot x_{n,r} & = & \nonumber  \\ x_{0,r} + \sum_{n\geq 1} x_{n,r}   
+   \sum_{i\geq 1}   \bar{f}_{\sigma}^i\cdot \kappa_s^i   \cdot    \left[\sum_{ n \geq i}    Q(i|n) \cdot x_{n,r} \right] 
\label{eq:evolving3} 
\end{eqnarray} 

Eq.~\ref{eq:widomSaft} simplifies to:
\begin{eqnarray}
\ln x_0 &=& \ln x_{0,r} - \ln\left[  1 + \sum_{i\geq 1} \bar{f}_{\sigma}^i \cdot \kappa_s^i \cdot    \left[\sum_{ n \geq i}    Q(i|n) \cdot x_{n,r} \right] \right ]  \nonumber \\ & =& \ln x_{0,r} + \ln X_0
\label{eq:widomSaft_gen}
\end{eqnarray}
Note that the summation $( \sum_{ n \geq i}    Q(i|n) \cdot x_{n,r} )$ contains all of the multi-body information in the reference fluid for a given patch geometry of the solute, at the density of the solution. This approach of representing the multi-body information can also be incorporated within SAFT and Wertheim's approach \cite{bansal_structure_2016} for different patch geometries.

Eqs.~\ref{eq:xn_asso} and \ref{eq:Xn} can be used to obtain the occupancy and bonding distributions, respectively, with 
$P(X_n|i)$ given by 
\begin{eqnarray}
P(X_i | n) = \frac {\bar{f}_{\sigma}^i \cdot \kappa_s^i \cdot Q( i | n)}{{K_n}/{K_{n,r}}} \, ,
\label{eq:PXn}
\end{eqnarray}
the bonded fractions in Eq.~\ref{eq:Xn} are
\begin{eqnarray}
X_i =  \bar{f}_{\sigma}^i \cdot \kappa_s^i \cdot X_0   \left[\sum_{ n \geq i}    Q(i | n) \cdot x_{n,r} \right]
\label{eq:Xn_gen}
\end{eqnarray}
As there is no association between the solvent molecules, the excess chemical potential of the solute due to association is reduced to
\begin{eqnarray}
\mu^{\rm ex}_{\rm asso}  = \ln X_0 \,   
\label{eq:brk_cpt1}
\end{eqnarray}
We consider several cases of solute patch geometry to test and illustrate the generality of the present approach.   Please note that as the patch size increases, as illustrated by Fig.\ref{fg:diff_patch}, we should expect multi-body interactions to become more important, making the analysis of association interactions more challenging. 
 
\section{Methods}\label{sc:methods}
 
 \subsection{Monte Carlo Simulations}
  
Monte Carlo simulations were carried out for the reference hard sphere systems and associating systems to validate the theory. The associating system contains a  single solute and 255 solvent particles \cite{bansal_structure_2016}. Solute-solvent association is allowed, but the solvent-solvent association is absent.   The system was equilibrated for 1 million steps with translational factors chosen to yield an acceptance rate of 0.3, and data was collected every 100 sweeps,  where a sweep is an attempted move of 
all the particles in the system. Analysis was carried out for different densities.
As discussed before \cite{bansal_structure_2016}, we use ensemble reweighting \cite{merchant_water_2011} to map $\{x_n\}$. 

 For associating systems, bonding ($X_n$) and occupancy ($x_n$) distributions were studied for a range of  critical angles for the solute and solvent molecules. 
The inner-shell radius (Fig.~\ref{fg:qcpdt}) is $r_c = 1.1d$, where $d$ is the diameter of the solute and solvent molecules. Since we are only exploring solute-solvent association in this work, for notational simplicity we dispense with the super-script (A or B) that classifies the patch according to its type. For association, 
 Table \ref{tab:diff_patch} gives the different solute patch sizes ($\theta_{c,\sigma}$) along with  the corresponding maximum bonding numbers ($N^{max}$), studied in this work.

 \begin{table}[htbp]
 	\centering
 	\caption{Patch size of the solute ($\theta_{c,\sigma}$) studied in the present work and the corresponding maximum bonding number ($N^{max}$). }
 	
 	\begin{tabular}{|c|ccccccc|}
 		\hline
 		\toprule
 	$\theta_{c,\sigma}$ & $27\degree$    & $35\degree$    & $40 \degree$     &$50 \degree$    & $60 \degree$     & $90 \degree$     & $180 \degree$  \\
 	$N^{max}$  & 1     & 2     & 3     & 4     & 5     & 7     & 12 \\
 	
 		\bottomrule
 		\hline
 	\end{tabular}%
 	\label{tab:diff_patch}%
 \end{table}%
 
  For most systems, a single bonding condition i.e. $ \theta_{c,s} = 27\degree$ was used for the solvent molecules. Some cases with  the same critical angle for the solute and solvent molecules, i.e. $\theta_{c,\sigma} =\theta_{c,s}$, were studied to compare QC results with the TPT2 framework developed earlier by Marshall et. al.\cite{marshall_thermodynamic_2014}.   For the spherically symmetric solute ($\theta_{c,\sigma} =180\degree$), the excess chemical potential of coupling the solute
 with solvent was also calculated using thermodynamic integration using a three-point Gauss Legendre quadrature rule \cite{Hummer:jcp96,bansal_structure_2016}.
 We analyzed cases of different densities and different association strengths between the solute and solvent molecules. 
 
 \subsection{Calculation of $Q(i | n)$}
 
Recall that  $Q( i | n ) = \langle \chi(i) | R^n \rangle_{r}$ is the probability that for $n$-solvent particles in the inner-shell, at least $i$ of those occupy the bonding patch and 
are available to bond with the solute.  For a defined solute patch geometry and occupancy $n$, we first obtain a viable $n$-solvent structure by adapting a method
that we formerly used to compute the $n$-solvent cluster integral \cite{bansal_structure_2016}. Then we generate new configurations by a Monte Carlo procedure. 
The Monte Carlo moves comprise a radial displacement that is restricted to be within $[d,r_c]$ and an orientational move over the surface of the sphere. The latter is performed by first picking at random one of three orthogonal axes, then choosing at random an angle by which to move
around that axis. The maximum angular move is adjusted to target 30 percent acceptance. After every 100$^{th}$ sweep, we analyze the configuration to compose  $Q( i | n )$.  We discretize the spherical polar coordinates $\cos \theta \in [-1, 1]$ and $\phi \in [0, 2\pi]$ into a grid of dimension $400\times 8000$. Then we sweep through the grid treating each grid site as a candidate direction for the bonding patch (Fig.~\ref{fg:diff_patch}). As we sweep through the grid sites, we collect statistics on 
how many of the $n$ particles in the cluster occupy the patch. From this information, we construct $Q(i|n)$. 

\section{Results and Discussions}\label{sc:results}
As discussed above, the present approach is general and can be applied to any geometry of the patch on the solute molecule. The factors $Q(i|n)$ discussed above play an important role in the theory. As we will also show below, a simple representation of $Q(i|n)$ can be obtained for the solute patch geometries that allow
the separation of orientation dependence from occupancy, as happens for a patch that can bond only once or a spherically symmetric patch. For all other cases, 
we need to explicitly calculate $Q(i|n)$. Appendix ~\ref{app:Qn} gives the values of $Q(i|n)$ for different patch geometries studied in this work. Throughout, we compare our results with Monte Carlo simulations, and where possible, with available SAFT models. 

\subsection{Patch geometries for the solute}

\begin{figure}[h!]
	\begin{center}
		\includegraphics[scale=0.85]{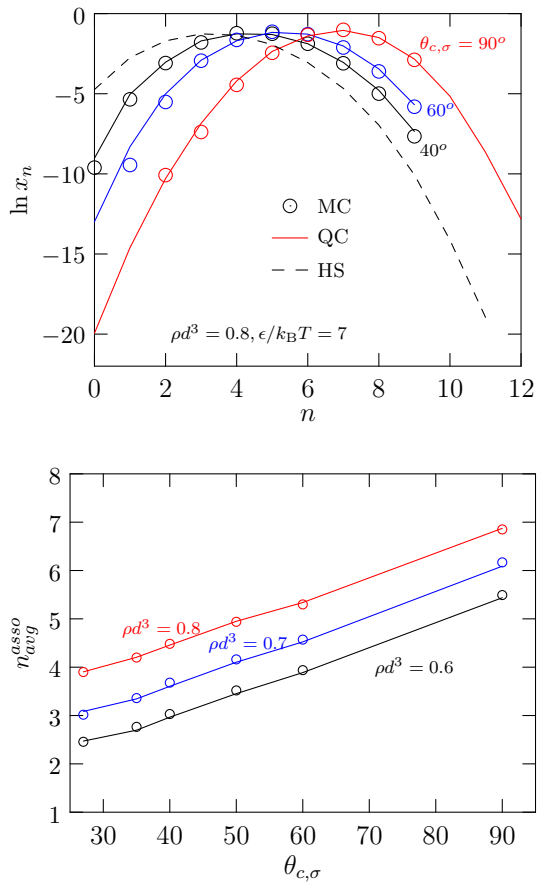}
	\end{center}
	\caption{Top panel: Comparison of the occupancy distribution for different patch geometries on the solute. The hard sphere reference distribution is also 
	shown using black dashed line. Bottom panel: Average occupancy for an associating solute for a range of patch geometries and for different reduced densities. The solvent molecules can bond only once to the solute molecule ($\theta_{c,s}=27 \degree $) and $\epsilon=7k_{\rm B}T$. Solid lines are QC predictions and symbols are Monte Carlo simulation results. \label{fg:occupancy_diff_patch}}
\end{figure}

We first consider a case where the solute can bond between 2 to 7 particles (Table~\ref{tab:diff_patch}), i.e.\ $\theta_{c,\sigma} \in [35, 90]$, and the solvent bonds only once, $\theta_{(c,s)} =27\degree$. Fig.~\ref{fg:occupancy_diff_patch} (Top) shows the occupancy distribution for different patch geometries of the solute for a reduced density of 0.8. Observe that when the patch size is increased from $35\degree$ to $90\degree$, the distribution moves towards higher occupancy states. The hard sphere distribution  (black dashed line) is also included to contrast with the cases including association. Fig.~\ref{fg:occupancy_diff_patch} (Bottom panel),
shows the average occupancy for a range of patch sizes on the solute and for different reduced densities of the solvent. We find that the QC based approach
is able to capture accurately the average occupancy for the entire range of densities.

\begin{figure*}[ht!]
	\begin{center}
		\includegraphics[scale=0.65]{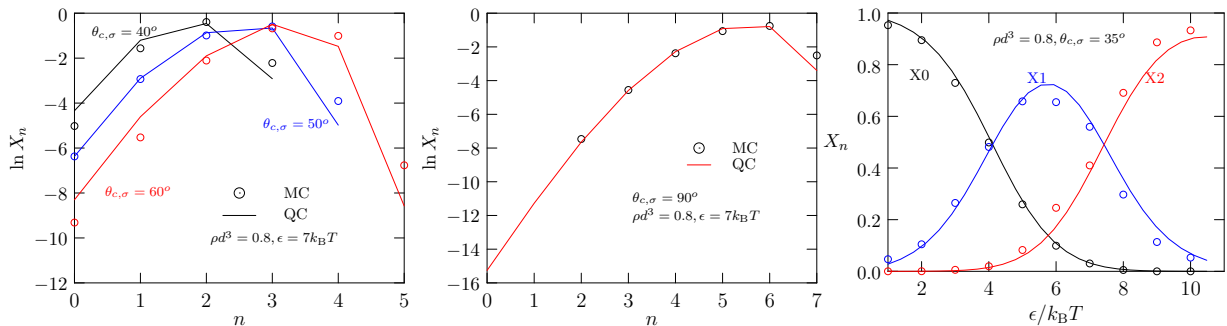}
	\end{center}
	\caption{Comparison of the bonding distribution for a solute with $40\degree$, $50\degree$, $60\degree$ (Left panel),  and $90\degree$ patch (Middle panel). Observe that the bonding fraction for the highest bonding state is invariably less well described by the theory. Right panel: Variation of different bonding fractions with association strength for a patch size of $35\degree$ at a density of 0.8. Rest as in Fig.~\ref{fg:occupancy_diff_patch}. \label{fg:bonding_35_40}}
\end{figure*}

Fig.~\ref{fg:bonding_35_40} shows the bonding distribution for the solute patch sizes of $35\degree$ to $90\degree$. The QC theory is 
able to  describe adequately the bonding distribution for all these patch sizes. Deviations for the highest bonded state were observed and are ultimately traced to numerical limitations  in estimating $Q(i|n)$ accurately. For high association strengths (such as $\epsilon=7k_{\rm B}T$), due to the higher powers of  Mayer $f$ function (Eq.~\ref{eq:PXn}), even slight errors in the magnitude of $Q(i|n)$ will have considerable impact on the bonding state. 

Fig.~\ref{fg:bonding_35_40} (Right panel) shows the variation in the bonding fractions with association strength for the case with $\theta_{c,\sigma}=35\degree$. As the
solute-solvent association strength is increased, the fraction of times the solute is non-bonded, the monomer fraction ($X_0$), decreases. The fraction
 bonded once ($X_1$) first increases and then decreases after a certain association threshold that starts favoring the double bonded fractions. This variation highlights the competition between the entropic effects that would favor the reference state, and the energetic effect, which favors higher bonding, of association. The QC theory is able to capture this behavior rather well. 

The decrease in the value of $X_1$ after a certain association strength as shown in Fig.~\ref{fg:bonding_35_40} cannot be captured within the TPT1 framework as the doubly bonded fractions are  ignored. Marshall et.al.\ \cite{marshall_thermodynamic_2014} extended SAFT beyond first order perturbation to include the double bonding condition. Their approach was based on a second order perturbation, but  terms beyond second bonding were left out. They studied mixtures of solvent and solute having the same bonding angles and allowed only solute-solvent association. As is clear from Fig.~\ref{fg:bonding_35_40}, as the bond angle increases beyond $35\degree$, the solute can bond more than three times and the second order perturbation is also inadequate. 

\begin{figure}[h]
	\begin{center}
		\includegraphics[scale=0.85]{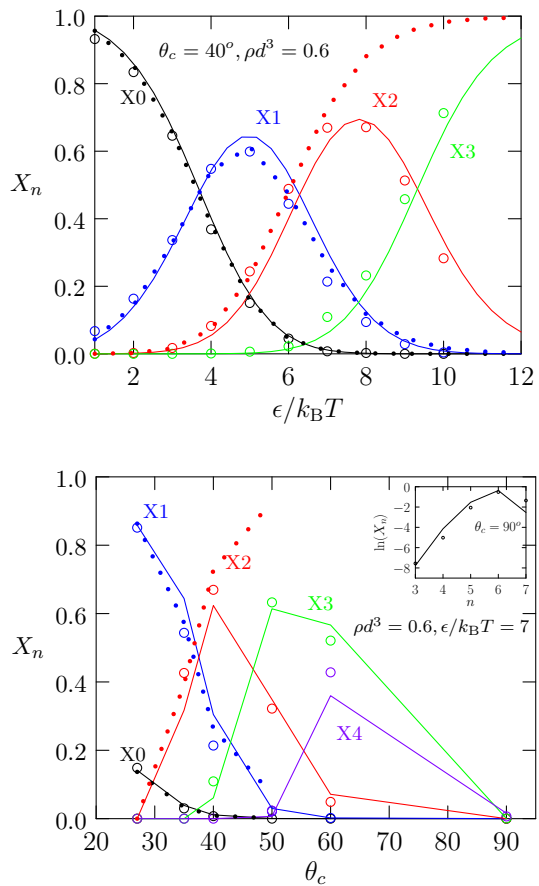}
	\end{center}
	\caption{Top:Comparison of the variation of different bonding fractions with the association strength, for a patch size of $40\degree$ at a density of 0.6.  
	Bottom: Comparison of bonding fractions for different patch sizes at a density of 0.6 and association strength of $\epsilon=7k_{\rm B}T$. For clarity, the bonding distribution for a $90\degree$ patch are shown in the inset figure in the bottom panel. The solvent molecules and solute molecule have same patch size ($\theta_{c,s}=\theta_{c,\sigma}=\theta_c $). Solute is infinitely dilute. Dotted lines are based on the approach by Marshall et.al.\cite{marshall_thermodynamic_2014}, solid lines are QC predictions and symbols are Monte Carlo simulation results.  \label{fg:Ben_comp}}
\end{figure}

In Fig.~\ref{fg:Ben_comp} (top panel), we compare the results of Marshall et. al.\ \cite{marshall_thermodynamic_2014} for a $40\degree$ patch at a density of $\rho d^3=0.6$
and for different association strengths. We find that for lower association strengths ($\epsilon/k_{\rm B}T<6)$, the second order perturbation is able to capture the bonding fractions. When the association strength is increased, fractions that are bonded three times increase and become more dominant. Necessarily, the second order perturbation is not adequate for these cases. QC is able to capture accurately all the bonding fractions across the association energy range. Fig.~\ref{fg:Ben_comp} (bottom panel) shows the excellent agreement for higher bonding fractions as the patch size on the solute is increased to $90\degree$ to model a Janus particle (Fig.~\ref{fg:diff_patch}). 

\subsection{Simplifications --- Single bonding patch}

In Wertheim's theory \cite{wertheim_fluids_1984,wertheim_fluids_1984-1} and SAFT \cite{chapman_new_1990}, a single bonding condition is assumed for the association 
sites. Thus $Q(i | n )=0$ for $i > 1$. This assumption is valid only a patch size $\theta_c=27\degree$ or less (Fig.~\ref{fg:diff_patch}). For this case, bonding and occupancy can be separated, and in particular, we have
\begin{eqnarray}
Q(1|n)=\kappa_{\sigma} \cdot C_1^n
\end{eqnarray}

where $C_1^n$ is a combinatorial factor accounting for the freedom to choose $1$ solvent molecule from $n$ solvent molecules in the observation volume and $\kappa_{\sigma}=(1-\cos(\theta_{c,\sigma}))/2$ is the probability that a solute molecule is oriented such that it can bond with the solvent. The
equilibrium constant ratio (Eq.\ref{eq:Kn_ratio_Q_red}) simplifies to 
\begin{eqnarray}
\frac{K_n}{K_{n,r}} &=&  \left[1+ \bar{f}_{\sigma}\cdot \kappa_s \kappa_{\sigma} \cdot C_1^n  \right]\, \\ \nonumber 
&=& \left[1+  \bar{f}_{\sigma}\cdot \kappa \cdot n  \right] 
\label{eq:Kn_ratio_1site}
\end{eqnarray}
where $\kappa = \kappa_s \kappa_{\sigma}$. The monomer fraction is then 
\begin{eqnarray}
\ln X_0 &=& - \ln\left[  x_{0,r} + \sum_{n\geq 1}   \left[1+   \bar{f}_{\sigma} \cdot \kappa \cdot n \right] \cdot x_{n,r} \right ] \nonumber \\ & =& - \ln\left[  1 + \bar{f}_{\sigma} \cdot \kappa   \sum_{n\geq 1} n\cdot x_{n,r} \right ]
\label{eq:widomSaft2}
\end{eqnarray}
In our previous work \cite{bansal_structure_2016}, we identified the sum in Eq.~\ref{eq:widomSaft2} as
\begin{eqnarray}
{F^{(1)}} = \sum_{n\geq 1} {n} \cdot x_{n,r} = n_{avg}^{\rm{hs}}= \rho_s\int\limits_{v} {d{{\vec r}_1} {g_{r}}\left( {{{\vec r}_1} |0} \right)}
\label{eq:Fn1}
\end{eqnarray}
where $n_{avg}^{\rm {hs}}$ is the average occupancy in the hard sphere reference, $\rho_s$ is the density of the solvent and $g_{r}$ is the pair correlation in the reference hard sphere system. Substituting the pair correlation form for the summation in Eq.~\ref{eq:widomSaft2}, we recover the SAFT representation \cite{chapman_new_1990} for the monomer fraction
\begin{eqnarray}
X_0 &=&  \left[  1 + \bar{f}_{\sigma} \cdot \kappa  \rho_s\int\limits_{v} {d{{\vec r}_1} {g_{r}}\left( {{{\vec r}_1} |0} \right)}  \right ]^{-1} \, , 
\label{eq:widomSaft_comp}
\end{eqnarray}
and the excess chemical potential of the solute due to association is 
\begin{eqnarray}
\mu^{\rm ex}_{\rm asso}  = \ln X_0 \, , 
\label{eq:cpt_site1}
\end{eqnarray}
again a well-known result within SAFT. 

\begin{figure*}[ht!]
	\begin{center}
		\includegraphics[scale=0.65]{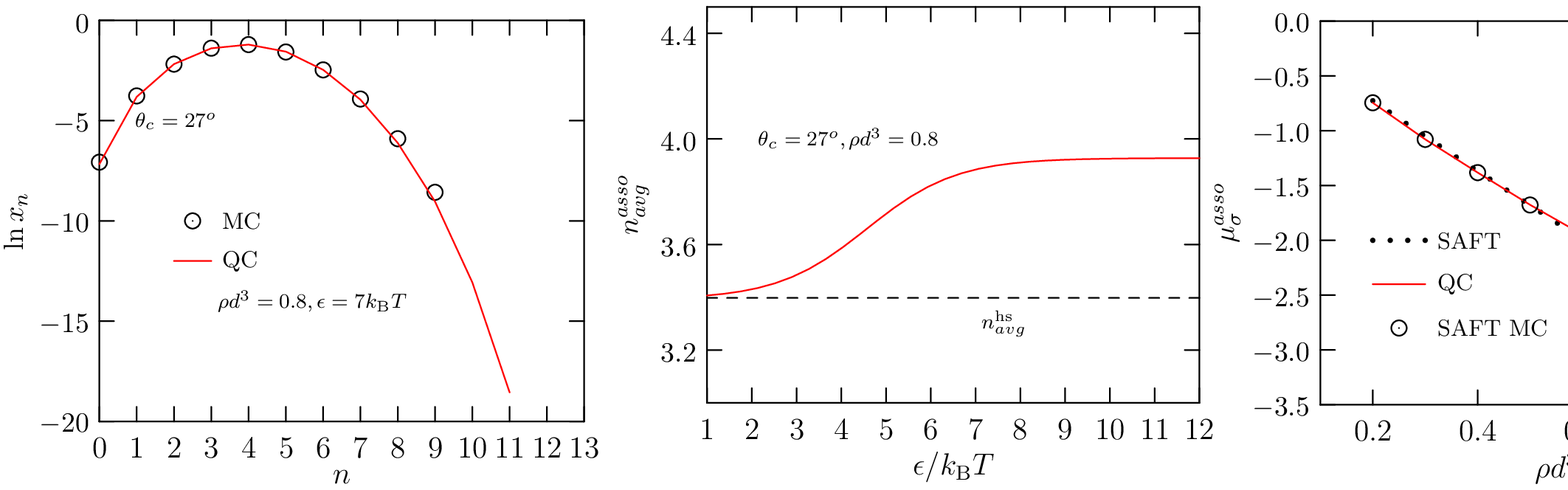}
	\end{center}
	\caption{Comparison of the occupancy distribution (Left) and chemical potential (Right) of an infinitely dilute solute with single bonding condition($\theta_{c,\sigma}=27 \degree $) for solute. Middle panel shows the average occupancy of associating solute for different association strengths. \label{fg:res_single_site}}
\end{figure*}
Fig.~\ref{fg:res_single_site} (Right) compares SAFT \cite{chapman_new_1990} and QC estimates for the residual chemical potential of the solute due to association. 
At higher densities, the SAFT approach is in error, whereas the QC approach describes the simulation results accurately. The deviation in the SAFT-based prediction
ultimately arises from the approximation 
 \begin{equation}
 r^2\cdot g_{r}(r)\approx d^2\cdot g_{r}(d)
 \label{eq:SAFT_approx}
 \end{equation}
 that is made within SAFT.  It was observed \cite{Chapman1988Cornell,marshall_resummed_2014} that this approximation is most accurate at a reduced density of 0.5. This approximation under-predicts bonding at the lower densities and over-predicts bonding at the higher densities, leading to a much too negative chemical potential
 at higher densities. We observe that, when accurate $n_{avg}^{\rm{hs}}$  values are provided in SAFT from Monte Carlo simulations \cite{bansal_structure_2016} (represented by SAFT MC in Fig.~\ref{fg:res_single_site} (Right)), excellent agreement with the QC estimate is obtained.  
 
The occupancy distribution can be obtained from Eq.~\ref{eq:xn_asso} by using equilibrium constant ratio from Eq.~\ref{eq:Kn_ratio_1site}
\begin{eqnarray}
x_n &=&  \frac {K_n}{K_{n,r}}  \cdot x_{n,r}\cdot X_0
\label{eq:xn_asso_1site}
\end{eqnarray}
For a reduced density of 0.8 and an association strength of $7 k_{\rm B}T$, Fig.~\ref{fg:res_single_site} (Left) shows that excellent agreement with Monte Carlo results is obtained with the QC theory for the entire distribution. The average occupancy for an associating solute can be obtained from the above distribution
 \begin{eqnarray}
 n_{avg}^{asso}&=&\sum_{n \ge 1} n\cdot x_n  \\ \nonumber
 &=& n_{avg}^{\rm {hs}}+\Delta n_{asso}
 \label{eq:xn_asso_1site}
 \end{eqnarray}
 where $n_{avg}^{\rm {hs}}=\sum_{n \ge 1} n\cdot x_{n,r}$.
Fig.~\ref{fg:res_single_site} (Middle panel) gives the results for a single bonding solute for different association strengths.

\subsection{Simplifications --- Spherically symmetric patch}

When the solute has a spherically symmetric patch $\theta_{c,\sigma}=180\degree$ (Fig.~\ref{fg:diff_patch}), all the orientations of the solute are favorable for bonding and hence the isolated cluster probabilities $Q(i|n)$, reduce to just choosing $i$ solvent molecules from the $n$ solvent molecules in the cluster, i.e.\ $Q(i|n)=C^n_i$.
\begin{eqnarray}
Q(i|n)=C^n_i = {{n}\choose{i}}~,  \forall~ i\leq n
\label{eq:Qn_sph}
\end{eqnarray}
Substituting Eq.~\ref{eq:Qn_sph} in Eq.~\ref{eq:Kn_ratio_Q_red}, we have 
\begin{eqnarray}
\frac{K_n}{K_{n,r}} =  \left[1+ \sum_{1\leq i \leq n} \bar{f}_{\sigma}^i\cdot \kappa_s^i \cdot {{n}\choose{i}}  \right]\, 
\label{eq:Kn_ratio_sph_sym}
\end{eqnarray}
The chemical contribution is obtained as

\begin{eqnarray}
\ln x_0 &=& \ln x_{0,r} - \ln\left[ 1   +\sum_{n \geq 1}  \bar{f}_{\sigma}^n\cdot \kappa_s^n  \cdot  \sum_{i\geq n} {{i}\choose{n}} \cdot x_{i,r} \right ] \nonumber \\ & =& \ln x_{0,r} + \ln X_0
\label{eq:widomSaft_SPh_sym}
\end{eqnarray}
%We can write

Recognizing that in the reference fluid, the average number of clusters with $n$ solvent molecules in the observation volume \cite{bansal_structure_2016,bansal_thermodynamics_2017,reiss_statistical_1959} is given by 
\begin{eqnarray}
{F^{(n)}} & =&  \sum_{i\geq n} {{i}\choose{n}}\cdot x_{i,r} \nonumber \\ & = & \frac{{{\rho_s^n}}}{{n!}}\int\limits_{v} {d{{\vec r}_1} \cdots \int\limits_{v} d{{\vec r}_n}{g_{r}}\left( {{{\vec r}_1} \cdots {{\vec r}_n}|0} \right)} \, ,
\label{eq:Fn}
\end{eqnarray}
we find that monomer fraction is 
\begin{equation}
X_0 =  \left[1 + \sum_{n\geq 1}  \bar{f}_{\sigma}^n \cdot \kappa_s^n   \cdot F^{(n)} \right]^{-1} \, 
\label{eq:X0_xph_sym}
\end{equation}
and the chemical potential of the solute is
\begin{eqnarray}
\mu^{\rm ex}_{\rm asso}  = \ln X_0 = -\ln \left[1 + \sum_{n\geq 1}  \bar{f}_{\sigma}^n \cdot \kappa_s^n   \cdot F^{(n)} \right]\, , 
\label{eq:mu_asso_sph}
\end{eqnarray}
expressions that were derived earlier within SAFT using the complete reference approach \cite{bansal_structure_2016,bansal_thermodynamics_2017}.

\begin{figure*}[!ht]
	\begin{center}
		\includegraphics[scale=0.65]{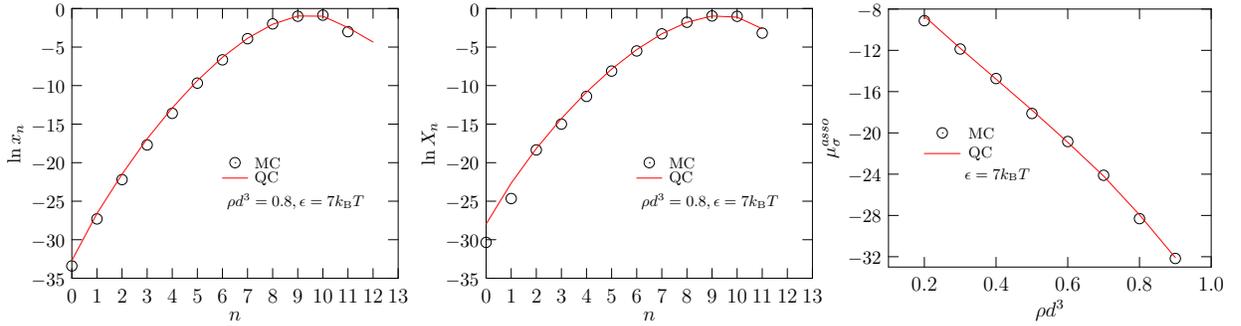}
	\end{center}
	\caption{Comparison of the occupancy distribution (Left), bonding fractions(Middle) and chemical potential (Right) of an infinitely dilute solute with spherically symmetric association ($\theta_{c,\sigma}=180 \degree $) \label{fg:res_sph_sym}}
\end{figure*}

Fig.\ \ref{fg:res_sph_sym} shows the comparison with Monte Carlo simulation for the occupancy distribution (Eq.~\ref{eq:xn_asso}), bonding distribution (Eq.~\ref{eq:Xn}), and the chemical potential(Eq.~\ref{eq:mu_asso_sph}) of an infinitely dilute solute with spherically symmetric association. The QC theory is able to capture the distribution
accurately for a high density and high association strength where multi-body correlations are important \cite{bansal_structure_2016}. 

\section{Conclusions}
  
The quasichemical (QC) approach offers a physically transparent and intuitive way to model the physics of association given the properties of a reference fluid. In particular, 
the approach provides a simple path to incorporate the physics of multi-body correlations. In the context of such multi-body correlations, the QC approach leads to the identification of a term, $Q(i|n)$,  that plays a central role in the theory. $Q(i|n)$ is a reference fluid property and is the conditional probability of $i$ particles being over the association patch 
given that $n$ particles are in the observation volume.  For a generic bonding patch we develop a Monte Carlo procedure for estimating $Q(i|n)$; 
for limiting cases where the orientation of the solvent can be decoupled from occupancy in the solute's inner-shell, we develop analytical expressions for $Q(i|n)$. 

In this work, for simplicity we study a mixture where solvent-solvent bonding is absent but the solvent can bond with the solute. Further, the solute is infinitely dilute. 
For such a mixture and for solutes with varying patch sizes, ranging from a solute that bonds only once, to solutes with  larger patch sizes including a Janus particle, and a particle whose entire surface is available for bonding, the theory leads to  predictions of bonding and occupancy that agree very well with results based on 
particle simulations. The quasichemical approach directly provides the excess chemical potential of the solute, which is expressed in terms of three contributions: the work to create a cavity the size of the solute's inner-solvation shell to accommodate the solute, the long-range work to couple the solute with the solvent when the inner-shell is empty of solvent, and the work to allow short-range association within the inner-shell. By construction, in the present study, the first two contributions are zero and the chemistry contribution simplifies to the logarithm of the monomer fraction of the solute molecule. For limiting cases, the expression for the nonbonded fraction of the solute is the same as the expression based on Wertheim's theory/SAFT. 

We note that the restriction of no solvent-solvent association is easily relaxed, potentially allowing us to model more complex mixtures. Further, our work hints at the possibility of readily modeling multi-body association within Wertheim/SAFT by using the $Q(i|n)$ factors identified by quasichemical theory. Results from these investigations will be presented later. 
   
 \section{Acknowledgment} 
 We acknowledge the Robert A. Welch Foundation (C-1241) and the Abu Dhabi National Oil Company (ADNOC) for support.
 
   \section{Appendix}\label{sc:appen}
   \subsection{Equilibrium constant}
   The equilibrium constant for the associating system can be expressed as \cite{bansal_thermodynamics_2017}
   \begin{equation}
   K_n = \frac{(e^{ \beta\mu^{\rm ex}_{s} })^n}{n!} e^{-\beta w(R^n)} \int\limits_vd{\vec r}_1\ldots\int\limits_v d{\vec r}_n \; e^{-\beta U_{\sigma s_n}(R^n)} \, ,
   \label{eq:kn1}
   \end{equation}
   and for the hard sphere reference as
   \begin{equation}
   K_{n,r} = \frac{(e^{ \beta\mu^{\rm hs}_{s} })^n}{n!} e^{-\beta w_{\rm hs}(R^n)} \int\limits_vd{\vec r}_1\ldots\int\limits_v d{\vec r}_n \; e^{-\beta U^{\rm hs}_{\sigma s_n}(R^n)} \, ,
   \label{eq:knr1}
   \end{equation}
   Taking ratio of the above two equations, we get
   \begin{widetext}
   \begin{eqnarray}
   \frac{K_n}{K_{n,r}} = e^{n\cdot \beta\Delta \mu^{\rm ex}_s} \cdot e^{\beta\Delta w(R^n)} \cdot \frac{\int\limits_vd{\vec r}_1\ldots\int\limits_v d{\vec r}_n \; e^{-\beta U_{\sigma s_n}(R^n)}}{\int\limits_vd{\vec r}_1\ldots\int\limits_v d{\vec r}_n \; e^{-\beta U^{\rm hs}_{\sigma s_n}(R^n)}} \, 
   \label{eq:Knratio1}
   \end{eqnarray}
   \end{widetext}
   we define
   \begin{eqnarray}
   Z_0 = \int\limits_vd{\vec r}_1\ldots\int\limits_v d{\vec r}_n \; e^{-\beta U^{\rm hs}_{\sigma s_n}(R^n)} \, 
   \label{eq:Knratio2}
   \end{eqnarray}
  Factoring $e^{-\beta  U^{\rm hs}_{\sigma s_n}(R^n)}$, we get
  \begin{widetext}
   \begin{eqnarray}
   \frac{K_n}{K_{n,r}} = e^{n\cdot \beta\Delta \mu^{\rm ex}_s} \cdot e^{\beta\Delta w(R^n)} \cdot \frac{\int\limits_vd{\vec r}_1\ldots\int\limits_v d{\vec r}_n \; e^{-\beta \Delta \tilde{U}_{\sigma s_n}(R^n)}e^{-\beta  U^{\rm hs}_{\sigma s_n}(R^n)}}{Z_0} \, 
   \label{eq:Knratio3}
   \end{eqnarray}
   \end{widetext}
  which reduces to Eq.~\ref{eq:Knratio}.

 \subsection{$Q(i|n)$ - Conditional isolated cluster probabilities}\label{app:Qn}
 As explained in section ~\ref{sc:methods}, we calculated $Q(i|n)$ for different patch sizes ($\theta_{(c,\sigma)}$) of the solute molecule. The values for different patch sizes studied in this work are presented in Tables \ref{tab:35_40_50} and \ref{tab:60_90}. It should be noted that for all the patch sizes, the maximum bonding numbers $N^{max}$ (see Table ~\ref{tab:diff_patch}) also give the maximum number that can be occupied in the patch region and hence,  $Q(i|n)=0, ~ \forall ~ i>N^{max}$. For different patch sizes, $Q(1|1)$ can be physically interpreted as the probability that the solvent molecule is in the patch region of the solute. The numerical values of $Q(1|1)$ obtained in this work agree with $k_{\sigma}=\left[1-\cos(\theta_{c,\sigma})\right]/2$, the probability that a solute molecule is oriented such that it can bond with the solvent.

     % Table generated by Excel2LaTeX from sheet '50tex'
     \begin{table*}[htbp]
     	\centering
     	\caption{Conditional isolated cluster probabilities for the critical angle on the solute, $\theta_{(c,\sigma)}=35^o,40^o$ and $50^o$.}
     	\begin{tabular}{|c|cc|c|ccc|r|cccr|}
     			\hline
     		\toprule
     		& \multicolumn{2}{c|}{$\theta_{(c,\sigma)}=35^o$} &       & \multicolumn{3}{c|}{$\theta_{(c,\sigma)}=40^o$} &       & \multicolumn{4}{c|}{$\theta_{(c,\sigma)}=50^o$} \\
     			\hline
     		\cmidrule{1-3}\cmidrule{5-7}\cmidrule{9-12}    n     & $Q(1|n)$ & $Q(2|n)$ &       & $Q(1|n)$ & $Q(2|n)$ & $Q(3|n)$ &       & $Q(1|n)$ & $Q(2|n)$ & $Q(3|n)$ & \multicolumn{1}{c|}{$Q(4|n)$ } \\
     			\hline
     		\cmidrule{1-3}\cmidrule{5-7}\cmidrule{9-12}    1     & 0.0904 & 0 &       & 0.1170 & 0 & 0     &       & 0.1786 & 0 & 0     & 0 \\
     		2     & 0.1804 & 0.0005 &       & 0.2324 & 0.0015 & 0     &       & 0.3453 & 0.0119 & 0     & 0 \\
     		3     & 0.2696 & 0.0017 &       & 0.3430 & 0.0080 & 7.81E-07 &       & 0.4981 & 0.0375 & 0.0002 & 0 \\
     		4     & 0.3580 & 0.0037 &       & 0.4541 & 0.0138 & 3.18E-05 &       & 0.6332 & 0.0802 & 0.0010 & 0 \\
     		5     & 0.4453 & 0.0069 &       & 0.5554 & 0.0294 & 8.95E-06 &       & 0.7495 & 0.1404 & 0.0032 & 3.29E-07 \\
     		6     & 0.5307 & 0.0118 &       & 0.6547 & 0.0471 & 0.0001 &       & 0.8435 & 0.2201 & 0.0080 & 2.52E-06 \\
     		7     & 0.6139 & 0.0190 &       & 0.7437 & 0.0748 & 0.0003 &       & 0.9126 & 0.3192 & 0.0185 & 1E-05 \\
     		8     & 0.6941 & 0.0292 &       & 0.8208 & 0.1141 & 0.0009 &       & 0.9595 & 0.4311 & 0.0383 & 3.73E-05 \\
     		9     & 0.7699 & 0.0439 &       & 0.8881 & 0.1628 & 0.0019 &       & 0.9856 & 0.5488 & 0.0729 & 0.0001 \\
     		10    & 0.8383 & 0.0659 &       & 0.9392 & 0.2263 & 0.0043 &       & 0.9966 & 0.6631 & 0.1259 & 0.0004 \\
     		11    & 0.8934 & 0.1011 &       & 0.9656 & 0.3099 & 0.0113 &       & 0.9974 & 0.7545 & 0.2120 & 0.0008 \\
     		12    & 0.9565 & 0.1284 &       & 0.9969 & 0.3840 & 0.0228 &       & 1.0000 & 0.8117 & 0.3313 & 0.0003 \\
     		\bottomrule
     			\hline
     	\end{tabular}%
     	\label{tab:35_40_50}%
     \end{table*}%

    % Table generated by Excel2LaTeX from sheet '50tex'
    \begin{table*}[ht]
    	\centering
    	\caption{Conditional isolated cluster probabilities for the critical angle on the solute, $\theta_{(c,\sigma)}=60^o$ and $90^o$.}
    	\begin{tabular}{|c|rrrrr|c|rrrrrrr|}
    		\toprule
    		\hline
    		& \multicolumn{5}{c|}{$\theta_{(c,\sigma)}=60^o$}              &       & \multicolumn{7}{c|}{$\theta_{(c,\sigma)}=90^o$} \\
    			\hline
    		\cmidrule{1-6}\cmidrule{8-14}    n     & \multicolumn{1}{c}{$Q(1|n)$} & \multicolumn{1}{c}{$Q(2|n)$} & \multicolumn{1}{c}{$Q(3|n)$} & \multicolumn{1}{c}{$Q(4|n)$} & \multicolumn{1}{c|}{$Q(5|n)$} &       & \multicolumn{1}{c}{$Q(1|n)$} & \multicolumn{1}{c}{$Q(2|n)$} & \multicolumn{1}{c}{$Q(3|n)$} & \multicolumn{1}{c}{$Q(4|n)$} & \multicolumn{1}{c}{$Q(5|n)$} & \multicolumn{1}{c}{$Q(6|n)$} & \multicolumn{1}{c|}{$Q(7|n)$} \\
    			\hline
    		\cmidrule{1-6}\cmidrule{8-14}    1     & 0.2500 & 0     & 0     & 0     & 0     &       & 0.5   & 0     & 0     & 0     & 0     & 0     & 0 \\
    		2     & 0.4667 & 0.0333 & 0     & 0     & 0     &       & 0.7925 & 0.2075 & 0     & 0     & 0     & 0     & 0 \\
    		3     & 0.6469 & 0.1013 & 0.0018 & 0     & 0     &       & 0.9329 & 0.5000 & 0.0671 & 0     & 0     & 0     & 0 \\
    		4     & 0.7850 & 0.2060 & 0.0090 & 1.3E-05 & 0     &       & 0.9850 & 0.7572 & 0.2428 & 0.0150 & 0     & 0     & 0 \\
    		5     & 0.8828 & 0.3395 & 0.0276 & 0.0001 & 0     &       & 0.9980 & 0.9170 & 0.5000 & 0.0830 & 0.002042 & 0     & 0 \\
    		6     & 0.9462 & 0.4895 & 0.0638 & 0.0005 & 5E-10 &       & 0.9999 & 0.9822 & 0.7567 & 0.2433 & 0.017769 & 9.39E-05 & 0 \\
    		7     & 0.9817 & 0.6415 & 0.1248 & 0.0020 & 3.34E-08 &       & 1     & 0.998112 & 0.9203 & 0.5000 & 0.079677 & 0.001888 & 0 \\
    		8     & 0.9951 & 0.7744 & 0.2245 & 0.0061 & 4.69E-07 &       & 1     & 0.999977 & 0.9846 & 0.7568 & 0.243163 & 0.015404 & 2.33E-05 \\
    		9     & 0.9993 & 0.8763 & 0.3582 & 0.0162 & 3.42E-06 &       & 1     & 1     & 0.9995 & 0.9231 & 0.5   & 0.076931 & 0.000497 \\
    		10    & 1.0000 & 0.9440 & 0.5140 & 0.0420 & 2.31E-05 &       & 1     & 1     & 1.0000 & 0.9928 & 0.756722 & 0.243278 & 0.007236 \\
    		11    & 1.0000 & 0.9821 & 0.6875 & 0.0802 & 0.0002 &       & 1     & 1     & 1     & 0.9994 & 0.951385 & 0.5   & 0.048615 \\
    		12    & 1.0000 & 0.9910 & 0.8745 & 0.1340 & 0.0005 &       & 1     & 1     & 1     & 1     & 0.992741 & 0.859332 & 0.140668 \\
    		\bottomrule
    			\hline
    	\end{tabular}%
    	\label{tab:60_90}%
    \end{table*}%
    
   \newpage 
%   \bibliography{references_QC} 

\begin{thebibliography}{25}
\expandafter\ifx\csname natexlab\endcsname\relax\def\natexlab#1{#1}\fi
\expandafter\ifx\csname bibnamefont\endcsname\relax
  \def\bibnamefont#1{#1}\fi
\expandafter\ifx\csname bibfnamefont\endcsname\relax
  \def\bibfnamefont#1{#1}\fi
\expandafter\ifx\csname citenamefont\endcsname\relax
  \def\citenamefont#1{#1}\fi
\expandafter\ifx\csname url\endcsname\relax
  \def\url#1{\texttt{#1}}\fi
\expandafter\ifx\csname urlprefix\endcsname\relax\def\urlprefix{URL }\fi
\providecommand{\bibinfo}[2]{#2}
\providecommand{\eprint}[2][]{\url{#2}}

\bibitem[{\citenamefont{Chapman et~al.}(1990)\citenamefont{Chapman, Gubbins,
  Jackson, and Radosz}}]{chapman_new_1990}
\bibinfo{author}{\bibfnamefont{W.~G.} \bibnamefont{Chapman}},
  \bibinfo{author}{\bibfnamefont{K.~E.} \bibnamefont{Gubbins}},
  \bibinfo{author}{\bibfnamefont{G.}~\bibnamefont{Jackson}}, \bibnamefont{and}
  \bibinfo{author}{\bibfnamefont{M.}~\bibnamefont{Radosz}},
  \bibinfo{journal}{Ind. Eng. Chem. Res.} \textbf{\bibinfo{volume}{29}},
  \bibinfo{pages}{1709} (\bibinfo{year}{1990}).

\bibitem[{\citenamefont{Chapman et~al.}(1988)\citenamefont{Chapman, Jackson,
  and Gubbins}}]{chapman_phase_1988}
\bibinfo{author}{\bibfnamefont{W.~G.} \bibnamefont{Chapman}},
  \bibinfo{author}{\bibfnamefont{G.}~\bibnamefont{Jackson}}, \bibnamefont{and}
  \bibinfo{author}{\bibfnamefont{K.~E.} \bibnamefont{Gubbins}},
  \bibinfo{journal}{Mol. Phys.} \textbf{\bibinfo{volume}{65}},
  \bibinfo{pages}{1057} (\bibinfo{year}{1988}).

\bibitem[{\citenamefont{Wertheim}(1984{\natexlab{a}})}]{wertheim_fluids_1984}
\bibinfo{author}{\bibfnamefont{M.~S.} \bibnamefont{Wertheim}},
  \bibinfo{journal}{J. Stat. Phys.} \textbf{\bibinfo{volume}{35}},
  \bibinfo{pages}{19} (\bibinfo{year}{1984}{\natexlab{a}}).

\bibitem[{\citenamefont{Wertheim}(1984{\natexlab{b}})}]{wertheim_fluids_1984-1}
\bibinfo{author}{\bibfnamefont{M.~S.} \bibnamefont{Wertheim}},
  \bibinfo{journal}{J. Stat. Phys.} \textbf{\bibinfo{volume}{35}},
  \bibinfo{pages}{35} (\bibinfo{year}{1984}{\natexlab{b}}).

\bibitem[{\citenamefont{Wertheim}(1986{\natexlab{a}})}]{wertheim_fluids_1986}
\bibinfo{author}{\bibfnamefont{M.~S.} \bibnamefont{Wertheim}},
  \bibinfo{journal}{J. Stat. Phys.} \textbf{\bibinfo{volume}{42}},
  \bibinfo{pages}{459} (\bibinfo{year}{1986}{\natexlab{a}}).

\bibitem[{\citenamefont{Wertheim}(1986{\natexlab{b}})}]{wertheim_fluids_1986-1}
\bibinfo{author}{\bibfnamefont{M.~S.} \bibnamefont{Wertheim}},
  \bibinfo{journal}{J. Stat. Phys.} \textbf{\bibinfo{volume}{42}},
  \bibinfo{pages}{477} (\bibinfo{year}{1986}{\natexlab{b}}).

\bibitem[{\citenamefont{Marshall et~al.}(2012)\citenamefont{Marshall, Ballal,
  and Chapman}}]{marshall_wertheims_2012}
\bibinfo{author}{\bibfnamefont{B.~D.} \bibnamefont{Marshall}},
  \bibinfo{author}{\bibfnamefont{D.}~\bibnamefont{Ballal}}, \bibnamefont{and}
  \bibinfo{author}{\bibfnamefont{W.~G.} \bibnamefont{Chapman}},
  \bibinfo{journal}{J. Chem. Phys.} \textbf{\bibinfo{volume}{137}},
  \bibinfo{pages}{104909} (\bibinfo{year}{2012}).

\bibitem[{\citenamefont{Marshall and
  Chapman}(2014)}]{marshall_thermodynamic_2014}
\bibinfo{author}{\bibfnamefont{B.~D.} \bibnamefont{Marshall}} \bibnamefont{and}
  \bibinfo{author}{\bibfnamefont{W.~G.} \bibnamefont{Chapman}},
  \bibinfo{journal}{Soft Matter} \textbf{\bibinfo{volume}{10}},
  \bibinfo{pages}{5168} (\bibinfo{year}{2014}).

\bibitem[{\citenamefont{Marshall and
  Chapman}(2013{\natexlab{a}})}]{marshall_molecular_2013}
\bibinfo{author}{\bibfnamefont{B.~D.} \bibnamefont{Marshall}} \bibnamefont{and}
  \bibinfo{author}{\bibfnamefont{W.~G.} \bibnamefont{Chapman}},
  \bibinfo{journal}{J. Chem. Phys.} \textbf{\bibinfo{volume}{139}},
  \bibinfo{pages}{104904} (\bibinfo{year}{2013}{\natexlab{a}}).

\bibitem[{\citenamefont{Marshall and
  Chapman}(2013{\natexlab{b}})}]{marshall_thermodynamic_2013}
\bibinfo{author}{\bibfnamefont{B.~D.} \bibnamefont{Marshall}} \bibnamefont{and}
  \bibinfo{author}{\bibfnamefont{W.~G.} \bibnamefont{Chapman}},
  \bibinfo{journal}{Soft Matter} \textbf{\bibinfo{volume}{9}},
  \bibinfo{pages}{11346} (\bibinfo{year}{2013}{\natexlab{b}}).

\bibitem[{\citenamefont{Paulaitis and Pratt}(2002)}]{lrp:apc02}
\bibinfo{author}{\bibfnamefont{M.~E.} \bibnamefont{Paulaitis}}
  \bibnamefont{and} \bibinfo{author}{\bibfnamefont{L.~R.} \bibnamefont{Pratt}},
  \bibinfo{journal}{Adv. Phys. Chem.} \textbf{\bibinfo{volume}{62}},
  \bibinfo{pages}{283} (\bibinfo{year}{2002}).

\bibitem[{\citenamefont{Beck et~al.}(2006)\citenamefont{Beck, Paulaitis, and
  Pratt}}]{lrp:book}
\bibinfo{author}{\bibfnamefont{T.~L.} \bibnamefont{Beck}},
  \bibinfo{author}{\bibfnamefont{M.~E.} \bibnamefont{Paulaitis}},
  \bibnamefont{and} \bibinfo{author}{\bibfnamefont{L.~R.} \bibnamefont{Pratt}},
  \emph{\bibinfo{title}{The Potential Distribution Theorem And Models Of
  Molecular Solutions}} (\bibinfo{publisher}{Cambridge University Press},
  \bibinfo{address}{Cambridge, UK}, \bibinfo{year}{2006}).

\bibitem[{\citenamefont{Pratt and Asthagiri}(2007)}]{lrp:cpms}
\bibinfo{author}{\bibfnamefont{L.~R.} \bibnamefont{Pratt}} \bibnamefont{and}
  \bibinfo{author}{\bibfnamefont{D.}~\bibnamefont{Asthagiri}}, in
  \emph{\bibinfo{booktitle}{Free Energy Calculations: {Theory} And Applications
  In Chemistry And Biology}}, edited by
  \bibinfo{editor}{\bibfnamefont{C.}~\bibnamefont{Chipot}} \bibnamefont{and}
  \bibinfo{editor}{\bibfnamefont{A.}~\bibnamefont{Pohorille}}
  (\bibinfo{publisher}{Springer}, \bibinfo{address}{Berlin, DE},
  \bibinfo{year}{2007}), vol.~\bibinfo{volume}{86} of
  \emph{\bibinfo{series}{Springer series in {Chemical Physics}}},
  chap.~\bibinfo{chapter}{9}, pp. \bibinfo{pages}{323--351}.

\bibitem[{\citenamefont{Bansal et~al.}(2016)\citenamefont{Bansal, Asthagiri,
  Cox, and Chapman}}]{bansal_structure_2016}
\bibinfo{author}{\bibfnamefont{A.}~\bibnamefont{Bansal}},
  \bibinfo{author}{\bibfnamefont{D.}~\bibnamefont{Asthagiri}},
  \bibinfo{author}{\bibfnamefont{K.~R.} \bibnamefont{Cox}}, \bibnamefont{and}
  \bibinfo{author}{\bibfnamefont{W.~G.} \bibnamefont{Chapman}},
  \bibinfo{journal}{J. Chem. Phys.} \textbf{\bibinfo{volume}{145}},
  \bibinfo{pages}{074904} (\bibinfo{year}{2016}).

\bibitem[{\citenamefont{Bansal et~al.}(2017)\citenamefont{Bansal,
  Valiya~Parambathu, Asthagiri, Cox, and Chapman}}]{bansal_thermodynamics_2017}
\bibinfo{author}{\bibfnamefont{A.}~\bibnamefont{Bansal}},
  \bibinfo{author}{\bibfnamefont{A.}~\bibnamefont{Valiya~Parambathu}},
  \bibinfo{author}{\bibfnamefont{D.}~\bibnamefont{Asthagiri}},
  \bibinfo{author}{\bibfnamefont{K.~R.} \bibnamefont{Cox}}, \bibnamefont{and}
  \bibinfo{author}{\bibfnamefont{W.~G.} \bibnamefont{Chapman}},
  \textbf{\bibinfo{volume}{146}}, \bibinfo{pages}{164904}
  (\bibinfo{year}{2017}).

\bibitem[{\citenamefont{Roh et~al.}(2005)\citenamefont{Roh, Martin, and
  Lahann}}]{roh_biphasic_2005}
\bibinfo{author}{\bibfnamefont{K.-H.} \bibnamefont{Roh}},
  \bibinfo{author}{\bibfnamefont{D.~C.} \bibnamefont{Martin}},
  \bibnamefont{and} \bibinfo{author}{\bibfnamefont{J.}~\bibnamefont{Lahann}},
  \bibinfo{journal}{Nature Materials; London} \textbf{\bibinfo{volume}{4}},
  \bibinfo{pages}{759} (\bibinfo{year}{2005}).

\bibitem[{\citenamefont{Walther and Müller}(2013)}]{walther_janus_2013}
\bibinfo{author}{\bibfnamefont{A.}~\bibnamefont{Walther}} \bibnamefont{and}
  \bibinfo{author}{\bibfnamefont{A.~H.~E.} \bibnamefont{Müller}},
  \bibinfo{journal}{Chem. Rev.} \textbf{\bibinfo{volume}{113}},
  \bibinfo{pages}{5194} (\bibinfo{year}{2013}).

\bibitem[{\citenamefont{Merchant and
  Asthagiri}(2009)}]{merchant_thermodynamically_2009}
\bibinfo{author}{\bibfnamefont{S.}~\bibnamefont{Merchant}} \bibnamefont{and}
  \bibinfo{author}{\bibfnamefont{D.}~\bibnamefont{Asthagiri}},
  \bibinfo{journal}{J. Chem. Phys.} \textbf{\bibinfo{volume}{130}},
  \bibinfo{pages}{195102} (\bibinfo{year}{2009}).

\bibitem[{\citenamefont{M{\"u}ller and
  Gubbins}(2001)}]{muller_molecular-based_2001}
\bibinfo{author}{\bibfnamefont{E.~A.} \bibnamefont{M{\"u}ller}}
  \bibnamefont{and} \bibinfo{author}{\bibfnamefont{K.~E.}
  \bibnamefont{Gubbins}}, \bibinfo{journal}{Ind. Eng. Chem. Res.}
  \textbf{\bibinfo{volume}{40}}, \bibinfo{pages}{2193} (\bibinfo{year}{2001}).

\bibitem[{\citenamefont{Bol}(1982)}]{bol_MontCarlo_1982}
\bibinfo{author}{\bibfnamefont{W.}~\bibnamefont{Bol}}, \bibinfo{journal}{Mol.
  Phys.} \textbf{\bibinfo{volume}{45}}, \bibinfo{pages}{605}
  (\bibinfo{year}{1982}).

\bibitem[{\citenamefont{Merchant et~al.}(2011)\citenamefont{Merchant, Shah, and
  Asthagiri}}]{merchant_water_2011}
\bibinfo{author}{\bibfnamefont{S.}~\bibnamefont{Merchant}},
  \bibinfo{author}{\bibfnamefont{J.~K.} \bibnamefont{Shah}}, \bibnamefont{and}
  \bibinfo{author}{\bibfnamefont{D.}~\bibnamefont{Asthagiri}},
  \bibinfo{journal}{J. Chem. Phys.} \textbf{\bibinfo{volume}{134}},
  \bibinfo{pages}{124514} (\bibinfo{year}{2011}).

\bibitem[{\citenamefont{Hummer and Szabo}(1996)}]{Hummer:jcp96}
\bibinfo{author}{\bibfnamefont{G.}~\bibnamefont{Hummer}} \bibnamefont{and}
  \bibinfo{author}{\bibfnamefont{A.}~\bibnamefont{Szabo}}, \bibinfo{journal}{J.
  Chem. Phys.} \textbf{\bibinfo{volume}{105}}, \bibinfo{pages}{2004}
  (\bibinfo{year}{1996}).

\bibitem[{\citenamefont{Chapman}(1988)}]{Chapman1988Cornell}
\bibinfo{author}{\bibfnamefont{W.~G.} \bibnamefont{Chapman}}, Ph.D. thesis,
  \bibinfo{school}{Cornell University} (\bibinfo{year}{1988}).

\bibitem[{\citenamefont{Marshall et~al.}(2014-04-28)\citenamefont{Marshall,
  Haghmoradi, and Chapman}}]{marshall_resummed_2014}
\bibinfo{author}{\bibfnamefont{B.~D.} \bibnamefont{Marshall}},
  \bibinfo{author}{\bibfnamefont{A.}~\bibnamefont{Haghmoradi}},
  \bibnamefont{and} \bibinfo{author}{\bibfnamefont{W.~G.}
  \bibnamefont{Chapman}}, \bibinfo{journal}{J. Chem. Phys.}
  \textbf{\bibinfo{volume}{140}}, \bibinfo{pages}{164101}
  (\bibinfo{year}{2014-04-28}).

\bibitem[{\citenamefont{Reiss et~al.}(1959)\citenamefont{Reiss, Frisch, and
  Lebowitz}}]{reiss_statistical_1959}
\bibinfo{author}{\bibfnamefont{H.}~\bibnamefont{Reiss}},
  \bibinfo{author}{\bibfnamefont{H.~L.} \bibnamefont{Frisch}},
  \bibnamefont{and} \bibinfo{author}{\bibfnamefont{J.~L.}
  \bibnamefont{Lebowitz}}, \bibinfo{journal}{J. Chem. Phys.}
  \textbf{\bibinfo{volume}{31}}, \bibinfo{pages}{369} (\bibinfo{year}{1959}).

\end{thebibliography}

 \end{document}